\documentclass{PoS}

\title{Supernova remnants in the very--high--energy sky: prospects for the Cherenkov Telescope Array}

\ShortTitle{Supernova remnants: prospects for CTA}

\author{\speaker{Pierre Cristofari}\\
        Columbia University\\
        E-mail: \email{pc2781@columbia.edu}}
        
\author{Roberta Zanin\\
	 Max-Planck-Institut fur Kernphysik\\
	}

\author{Stefano Gabici\\
	APC, AstroParticule et Cosmologie, Universit\'e Paris Diderot, CNRS/IN2P3, CEA/Irfu, Observatoire de Paris, Sorbonne Paris Cit\'e \\
	}

\author{Brian T. Humensky\\
	Physics Department, Columbia University\\
	}
\author{Marcos Santander\\
	Barnard College, Columbia University\\
	}
\author{Regis Terrier\\
	APC, AstroParticule et Cosmologie, Universit\'e Paris Diderot, CNRS/IN2P3, CEA/Irfu, Observatoire de Paris, Sorbonne Paris Cit\'e\\
	}
	\author{for the CTA consortium\\	}

\abstract{The Cherenkov Telescope Array is expected to lead to the detection of many new supernova remnants (SNRs) in the TeV and multi--TeV range. In addition to the individual study of each SNR, the study of these objects as a population can help constraining the parameters describing the acceleration of particles and increasing our understanding of the mechanisms involved. We present Monte Carlo simulations of the population of Galactic SNRs emitting TeV gamma rays. We also discuss how the simulated population can be confronted with future observations to provide a novel test for the SNR hypothesis of cosmic ray origins.}

\FullConference{35th International Cosmic Ray Conference – ICRC217-\\
		10-20 July, 2017\\
		Bexco, Busan, Korea}

\begin{document}

\section{Introduction}
The ever increasing number of supernova remnants  (SNRs) detected in the TeV range constitutes a major asset for the study of the acceleration mechanisms happening at strong shocks \cite{donath2016}. Indeed, TeV gamma rays are produced when very--high--energy particles interact with their environment, and the observation of excesses of TeV gamma rays at several astrophysical objects testify of efficient acceleration mechanisms. 
The case of SNRs is especially interesting, because they have been widely pointed as the most probable sources of Galactic cosmic rays (CRs) (i.e. at least up to energies of the \textit{knee} $\sim 1$ PeV)~\cite{bartoli2015}. The SNR hypothesis is supported by several strong arguments, such as energy considerations, showing that converting a fraction of the order of 10\% of the total explosion energy of supernovae into CRs  can explain the measured level of CRs at the Earth. Another strong argument is the diffusive shock acceleration mechanism, operating at SNR shocks, capable of explaining the power--law slope of accelerated particles, compatible with local measurements of CRs. 

However, these supporting arguments are not enough to make the SNR hypothesis a definitive answer to the question of the origin of Galactic CRs. CRs are mainly protons, and the sources of CRs are therefore expected to demonstrate that they can efficiently accelerate protons up to the \textit{knee}. The observation of gamma rays in the TeV range from SNRs attest of particle acceleration, but can often be explained by accelerated electrons as well as accelerated protons. Indeed, accelerated electrons can undergo inverse Compton scattering on soft photons of the CMB, and protons can interact with the interstellar medium to produce gamma rays through pion decay. In the TeV range, the situation is therefore often unclear, and motivates further testing of the SNR hypothesis. 

The actual population of SNRs in the TeV range comes from targeted observations and from systematic Galactic surveys. The case-by-case study of all these SNRs has greatly improved the understanding of the community on acceleration mechanisms, and the modeling efforts have in many cases allowed satisfying interpretation of the origin of their TeV emission. But the growing number of detections motivates a study of the entire population as such. 
In this context, it is possible to simulate the expected SNR population and compare it with actual observations from current TeV instruments, therefore providing a test for the SNR hypothesis. 
The demonstrated efficiency of systematic surveys and the perspective of deeper all--sky survey, such as the one proposed by CTA suggest the detection in the coming years of a SNR population significantly larger than the current one~\cite{CTA2013,cristofari2017}. This population, confronted with theoretical simulations will help test again the role of SNRs, and constrain the parameters governing particle acceleration at SNR shocks. 

 \section{Method}
 
We rely on Monte Carlo methods to simulate the population of SNRs potentially detectable by CTA. 
For repeated realizations ($10^3$), we simulate the time and location of supernova explosions in the Galaxy, assuming a rate of 3 SN per century, and a spatial distribution described as in~\cite{faucher}. Two mechanisms, via four types of progenitors, are considered:~thermonuclear (type Ia) and core--collapse (types Ib/c, IIP, IIb). The relative rates and typical parameters associated to each type, such as the total supernova explosion energy, the velocity of the wind and the mass of the ejecta, are adopted as in~\cite{seo}, so that every simulated supernova is assigned a type and corresponding parameters. 
At the location of each supernova, the typical value of the interstellar medium (ISM) is derived from surveys of atomic and molecular hydrogen~\cite{HI,H2}. The evolution of the shock radius $R_{\rm sh}$ and velocity $u_{\rm sh}$ is computed using analytical and semi--analytical description of~\cite{chevalier,pz05}. 

Finally, the gamma--ray luminosity of each SNR is computed. The contribution of protons and electrons is taken into account. At the shock, the particles are assumed to be accelerated with a slope following a power--law in momentum $n(p) \propto p^{-\alpha}$, where $\alpha$ is treated as a parameter in the range $4.1 - 4.4$. At the shock, we assumed that a fraction $\xi_{\rm CR}$ of the ram pressure of the shock expanding through the ISM is converted into CRs, where $\xi_{\rm CR} \approx 0.1$, and the shock compression factor is $\sigma=4$.  
The distribution of CRs inside the SNR is computed by solving a transport equation and the structure of the interior of the SNR is derived by solving the gas continuity equation, as in~\cite{pz03,pz05}.
The maximum momentum reached by protons is computed by assuming that protons escape the shock when their diffusion length equates a fraction $\zeta \approx 0.1$ of the shock radius, adopting a Bohm diffusion coefficient, this leads to $p_{\max} \propto R_{\rm sh} u_{\rm sh} B_{\rm down}$, where $B_{\rm down}$ is the magnetic field downstream of the shock. In order to account for magnetic field amplification downstream of the shock, and without making any assumption on the type of mechanism involved in the amplification, we describe $B_{\rm down} = \sigma B_{0}\sqrt{({u_{\rm sh}}/{v_{\rm d}})^{2}+1}$, where $v_{\rm d}$ is explicited in~\cite{zirakashviliaharonian2010} .
The hadronic contribution to the gamma--ray spectrum is then calculated following the approach of~\cite{kelner2006}, weighted by a factor 1.8 to take into account nuclei heavier than hydrogen. 

We follow by computing the leptonic component. 
The spectrum of electrons is parametrized adopting the same spectral shape as protons $\propto p^{-\alpha}$, weighted by a factor $K_{\rm ep}$, for momenta $p$ < $p_{\rm break}$, where $p_{break}$ accounts for radiative losses. above $p_{\rm break}$ the electron spectrum steepens by one order and follows $\propto p^{-\alpha-1}$~\cite{morlino2012}. The maximum momentum of electrons is reached when the synchrotron loss time is of the order of the acceleration rate. The gamma--ray luminosity from inverse Compton scattering of electrons on the cosmic microwave background is computed following the description proposed by~\cite{gould}.

The approach presented here is described more in detail in \cite{cristofari2013,cristofari2017} and was used to provide a statistical test for the SNR hypothesis of the origin of CRs. 

 \section{Results}
\begin{figure}
\centering
\includegraphics[width=.5\textwidth]{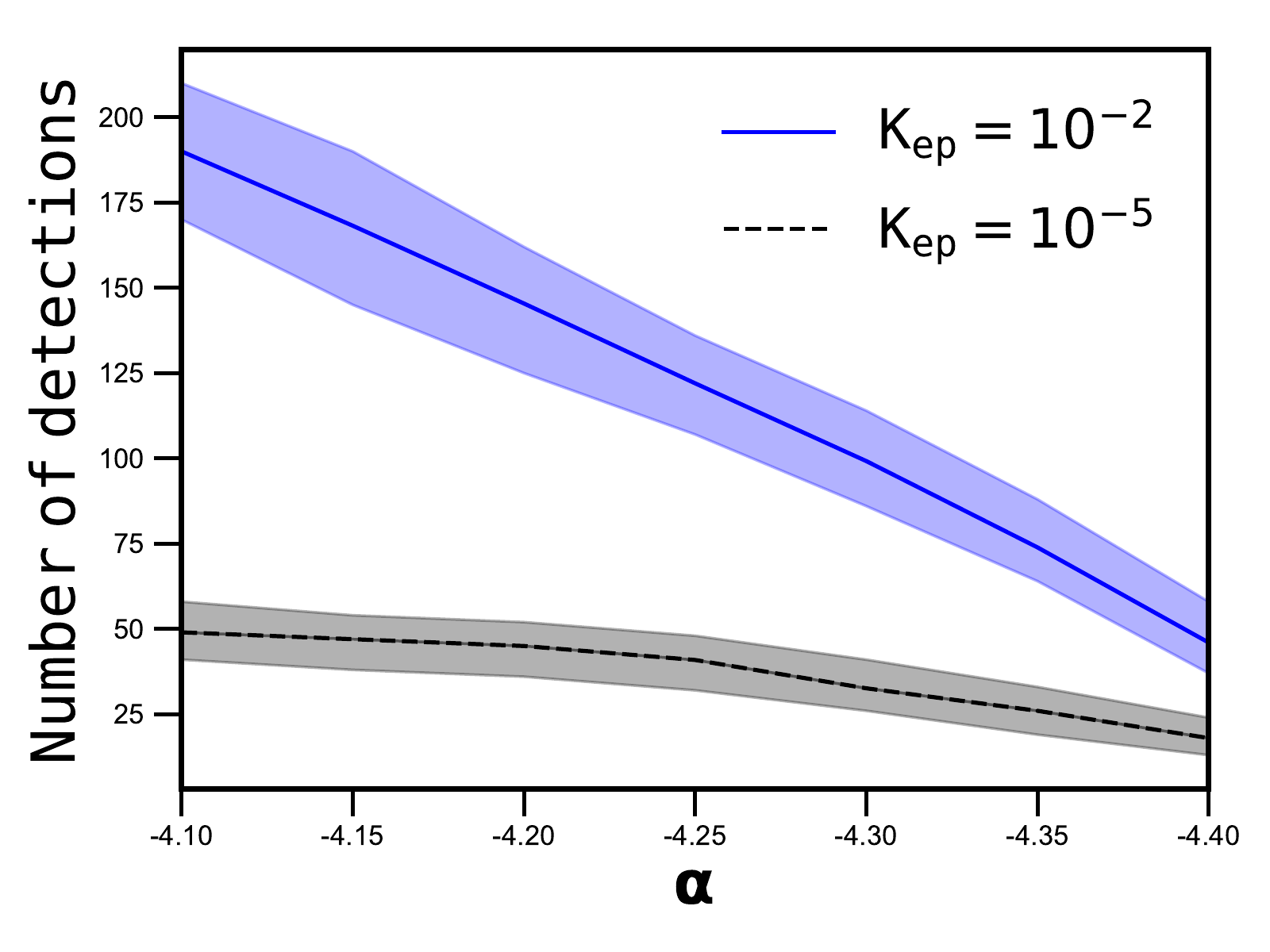}
\caption{SNRs in the simulated Galactic plane survey of CTA with integral gamma--ray flux F($>$~1TeV)~$\geq$~1~mCrab, as a function of the parameter $\alpha$. The blue (solid) and black (dashed) curve correspond respectively to $K_{\rm ep }= 10^{-2}$ and $K_{\rm ep }= 10^{-5}$. In each case the +/- standard deviation is shown.}
\label{fig:plot_alpha}
\end{figure}

 Two strategies have been proposed for the Galactic survey of CTA: an all--sky survey where a typical integrated sensitivity of $\approx 3$ mCrab could be reached, and a Galactic plane survey (GPS) centered on the Galactic center ($| l  | < 60^{\circ}$, $| b | < 2^{\circ}$) where a sensitivity of $\approx 1$ mCrab could be reached~\cite{CTA2013}. 
 Using the method described in the previous paragraph, we simulate the population that CTA could expect to detect in the GPS while performing an all--sky survey and plot in Fig.~\ref{fig:plot_alpha} the number of number of simulated SNRs with integral gamma--ray flux above F($>$~1TeV) greater than 1 mCrab. 
 In order to take into account the extension of the simulated SNRs, the sensitivity of CTA was degraded linearly by the sources apparent size when this becomes larger the typical point spread function of the instrument, i.e. $\approx 3$ arcmin at 1 TeV. 
 The range of parameters adopted for Fig.~\ref{fig:plot_alpha} of $\alpha=4.1-4.4$ and $K_{\rm ep}=10^{-2}-10^{-5}$ have been proposed by theoretical studies~\cite{vladimirdrift,ohira,stefanoescape,donRXJ,morlino2012}. 
The most \textit{optimistic} situation represented, where $\alpha=4.1$ and $K_{\rm ep}=10^{-2}$ lead to a number of $\approx 190^{+20}_{-20}$ potentially detectable SNRs. The most \textit{pessimistic} situation, where $\alpha=4.4$ and $K_{\rm ep}=10^{-5}$ lead to $\approx 18^{+6}_{-5}$ potential detection. 
Other effects should also be taken into account, such as for example the issue of source confusion: with the improved sensitivity, many of the new sources could overlap making the identification of SNRs problematic~\cite{dubus2013}. 

\begin{figure}
\centering
\includegraphics[width=.5\textwidth]{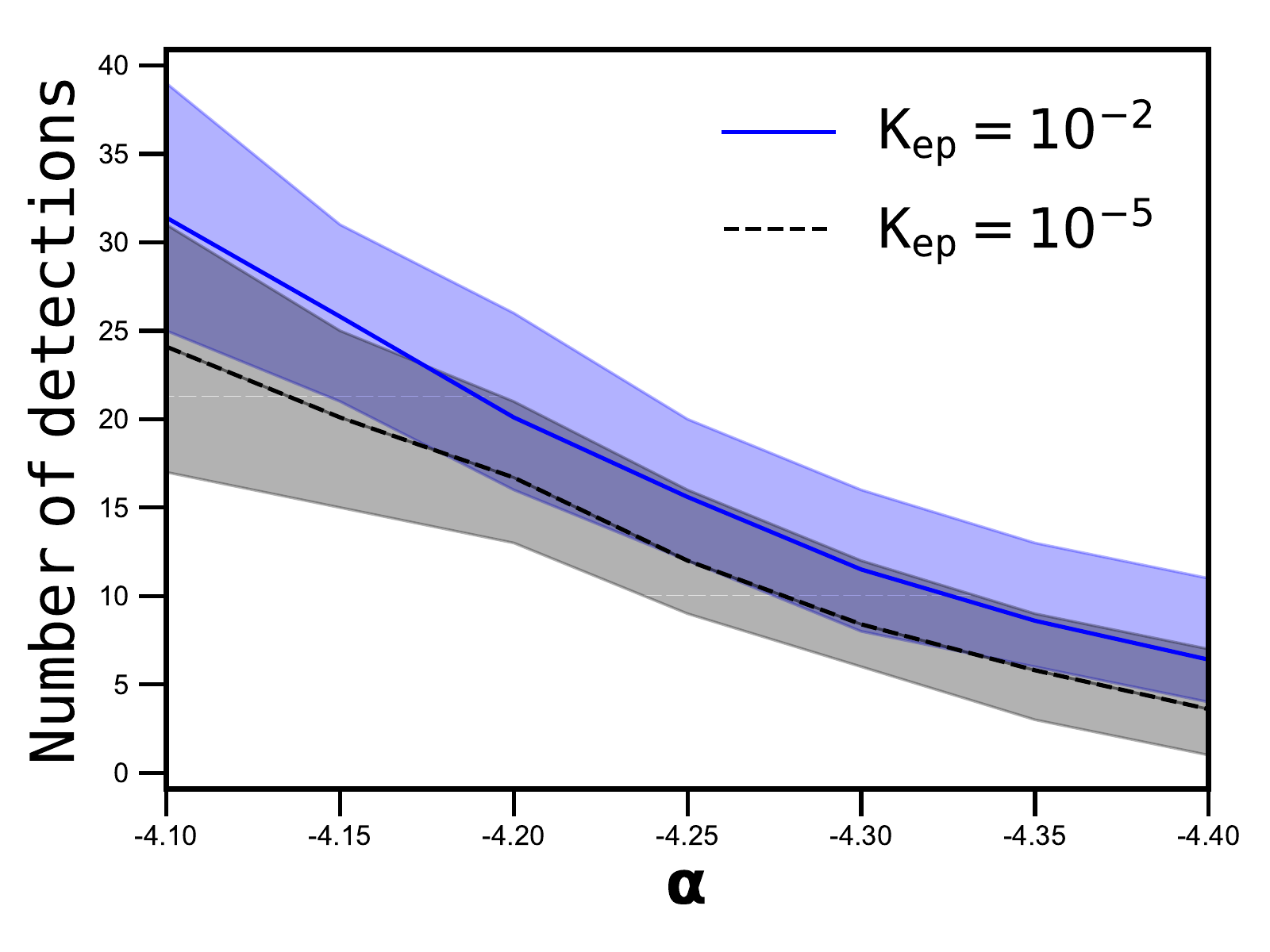}
\caption{Situation analogous to Fig.~\ref{fig:plot_alpha} with integral flux F($>$~10 TeV) $\geq$ 10 mCrab.}
\label{fig:plot_alpha10}
\end{figure}

The main take--away message is that the different sets of parameters can lead to remarkably different populations, and that in the TeV range, CTA should be able to further constrain these parameters. 
A more detailed description of the characteristics (age, distance, size) of the simulated populations, and a discussion on the robustness of our approach, can be  found in~\cite{cristofari2017}. 
Considering integral gamma--ray fluxes above 10 TeV and a sensitivity of 10 mCrab, the number of detection is $\approx 30^{+8}_{-7}$ and $\approx 4^{+2}_{-2}$ in the extreme situations. In Fig.~\ref{fig:plot_alpha10}, it thus becomes obvious that it will be more difficult to constrain the parameter $K_{\rm ep}$ at 10 TeV than at 1 TeV. 
This is expected, given the fact that in the multi--TeV range, the leptonic contribution to the gamma--ray emission becomes less important.

 \section{Conclusions}
 
 Next generation instruments operating in the TeV range, such as CTA, are expected to lead to many new SNR detections, especially thanks to systematic Galactic surveys. 
 In the most optimistic scenarios, the SNR population accessible by CTA could somewhat be comparable to the one detected at other wavelength, such as in the GHz range, where $\lesssim 300$ SNRs have been reported~\cite{green2015}.
More than predictions on what CTA should achieve, our work show that the parameters governing particle acceleration should lead to very different situations in terms of detection of SNR population, and that CTA  should be able to discriminate between these situations. This is not the case with the results of current TeV instruments, where the low number of detections can not at this stage efficiently been used for such an analysis. The population detected by CTA should therefore be confronted to our simulations in order to provide a novel consistency test of the SNR hypothesis, and improve our understanding of the role of SNRs in the acceleration of very--high--energy particles. 

\section*{Acknowledgements}
This work was conducted in the context of the CTA Consortium. The authors gratefully acknowledge financial support from the agencies and organizations listed here: http://www.cta-observatory.org/consortium\_acknowledgments, and thank the CTA consortium. SG acknowledges support from the Programme National Hautes Energies (CNRS). SG and RT acknowledge support from the Observatoire de Paris (Action Fédératrice Preparation à CTA). PC acknowledges support from the Columbia University Frontiers of Science fellowship.

\end{document}